\documentclass[letterpaper,12pt]{article}

\usepackage{mathptmx}

\usepackage{graphics}

\usepackage[authoryear,comma,sectionbib]{natbib}

\usepackage{amsthm,amsmath}
\usepackage[utf8]{inputenc} %unicode support
\usepackage{amsmath}
\usepackage{amsfonts}
\usepackage{algorithm,algorithmic}
\usepackage{comment}
\usepackage{tabularx,ragged2e,booktabs}
\usepackage{graphicx}
\usepackage{lscape}
\usepackage{color}
\usepackage{lipsum}
\usepackage{authblk}
\usepackage{parskip}

\RequirePackage{geometry}
\begin{document}

\title{Statistical development and assessment of summary measures to account for isotopic clustering of Fourier transform mass spectrometry data in clinical diagnostic studies}

%\begin{comment}
\centering
\author[1*]{Alexia Kakourou}
\author[2]{Werner Vach}
\author[3]{Simone Nicolardi}
\author[3]{Yuri van der Burgt}
\author[1]{Bart Mertens}

\affil[1]{Department of Medical Statistics and Bioinformatics, Leiden University Medical Center, Leiden, The Netherlands}
\affil[2]{Center for Medical Biometry and Medical Informatics, University of Freiburg, Freiburg, Germany}
\affil[3]{Center for Proteomics and Metabolomics, Leiden University Medical Center, Leiden, The Netherlands}

\renewcommand\Affilfont{\itshape\small}

\renewcommand\Authands{ and }
\date{}

\maketitle

%\thispagestyle{empty}
%\end{comment}

%% Do NOT include any fronmatter information; including the title, author names,
%% institutes, acknowledgments and title footnotes (author information, funding
%% sources, etc.). Start the document with the first section or paragraph of
%% the article.

\begin{comment}
\author{
  Alexia Kakourou,
  Werner Vach,
  Simone Nicolardi,\\
  Yuri van der Burgt and
  Bart Mertens
   \\
   \\
  Institution Institution Institution Institution Institution Institution Institution \\
  Institution Institution Institution Institution Institution Institution Institution
}
\end{comment}

%\MessageBreak\originalmaketitle\

\newpage
\abstract{Mass spectrometry based clinical proteomics has emerged as a powerful tool for high-throughput protein profiling and biomarker discovery. Recent improvements in mass spectrometry technology have boosted the potential of proteomic studies in biomedical research. However, the complexity of the proteomic expression introduces new statistical challenges in summarizing and analyzing the acquired data. Statistical methods for optimally processing proteomic data are currently a growing field of research. In this paper we present simple, yet appropriate methods to preprocess, summarize and analyze high-throughput MALDI-FTICR mass spectrometry data, collected in a case-control fashion, while dealing with the statistical challenges that accompany such data. The known statistical properties of the isotopic distribution of the peptide molecules are used to preprocess the spectra and translate the proteomic expression into a condensed data set. Information on either the intensity level or the shape of the identified isotopic clusters is used to derive summary measures on which diagnostic rules for disease status allocation will be based. Results indicate that both the shape of the identified isotopic clusters and the overall intensity level carry information on the class outcome and can be used to predict the presence or absence of the disease.
\\
\mbox{}
\\
KEY WORDS: Clinical mass-spectrometry based proteomics, Fourier transform mass spectrometry, Isotopic distribution, Prediction}

\newpage

\section{Introduction}
\indent Mass spectrometry (MS) based clinical proteomics has emerged as a powerful analytical tool for protein profiling of patient body fluids. It is widely used in cancer-associated marker discovery, with the aim to better understand the specific disease as well as to exploit diagnostic, prognostic and therapeutic potential. Proteomic methods can be used for the comparison of protein profiles, consisting of high-dimensional features (peaks), of which the presence or intensity can depend on the physiological and pathological condition of the individual. Examples are case-control studies for the construction of diagnostic rules or prognostic studies for the prediction of the disease outcome. Statistical analysis of data collected in the context of such studies offers the opportunity to improve detection ability and facilitate early diagnosis and prognosis.
\\
\indent Most often, peptide and protein analysis by MS occurs either through electrospray ionization (ESI) or via matrix-assisted laser desorption/ionization (MALDI). Whereas ESI yields multiply charged species, MALDI (predominantly) results in singly charged species. In this study MALDI mass spectra are considered. These spectra, or profiles, can be obtained via coupling with a time-of-flight mass analyzer (MALDI-TOF) or with a Fourier transform ion cyclotron resonance system (MALDI-FTICR). There are inherent challenges in characterizing protein expression levels with regard to sensitivity and reproducibility (\citealp{diamand,ander}). Different types of mass spectrometers vary in their ability to overcome these challenges. Fourier-transform (FT)-based technologies have shown to be provide powerful approaches for obtaining biomarker signatures. Particularly Fourier-transform MS-platforms are extremely powerful for the analysis of complex mixtures. The main advantages of using this technology, compared to TOF-MS, are related to 1) the ultrahigh mass resolving power (routinely more than 100,000) which allows the analysis of large proteins and complex mixtures, 2) the mass accuracy and precision which allow more reliable identification of the detected species and 3) the wide dynamic range which is favorable for the detection of low abundant components.
\\
\indent Although various approaches on preprocessing and interpreting high - resolution mass spectral data have been reported previously (\citealp{averagine,roc,palm,valk,yuri}), these have been used for the analysis of single spectra only, hence none of these has ever been evaluated in the context of clinical applications, particularly in the case of class calibration and prediction problems.
\\
\indent The main focus of this paper is to introduce an overview of methods to preprocess, summarize and analyze high-throughput MALDI-FTICR mass spectrometry data, collected in the context of a pancreatic cancer case-control study, while dealing with the statistical challenges which accompany this specific type of data. We propose a computationally simple and fast method to preprocess the raw spectra and translate the proteomic expression into a condensed data set. Our approach uses the fact that in a MALDI spectrum singly charged species are spaced 1 mass unit within each isotopic distribution. Using this property, the complete expression in the individual spectra can be reduced to clusters of isotopic expression on which summary measures can be defined. For the remainder of the paper, we will refer to the derived clusters as isotopic clusters, and to the expression within each isotopic cluster as isotopic cluster pattern. We investigate various ways of summarizing the observed isotopic cluster pattern with the ultimate goal to evaluate the different choices with respect to their predictive capacity. To derive the summary measures we use information on either the intensity or the shape of the observed isotopic cluster pattern. In this way we investigate whether there is additional information in the shape of the isotopic cluster patterns which can be used to predict the health status of future patients. We finally explore the potential of improving predictions by combining both intensity and shape.
\\
\indent The structure of the article is as follows: In Section 2 we will briefly overview the data and its properties. The procedure employed to preprocess the raw data is then presented, followed by an overview of the methods used to summarize the proteomic profiles. In Section 3 a comparative analysis of the proposed summary measures is presented in terms of their predictive ability. We assess the impact of the different choices on class calibration and prediction, by using the proposed summary measures as input variables into ridge regression, and by evaluating the predictive performance of the resulting discriminating rules. Moreover, we investigate the pure information of shape by isolating the shape information from the intensity information. Finally the predictive potential of diagnostic rules based on integrating both types of information will be exploited.

\section{Materials and Methods}

\subsection{Data description}
We consider a case-control study, the design of which is described in detail in \citet{simone}. The experiment involves a total of 273 individuals, consisting of 88 pancreatic cancer patients and 185 healthy volunteers. From each of the included individuals, a serum sample was collected, stored and processed according to a standardized protocol. The original study was set up to define a calibration set and a separate validation set, the samples of which were collected in a later period than the samples of the calibration set. More specifically, for the calibration set, serum samples were collected from 49 pancreatic cancer patients prior to surgery and 110 healthy controls (age- and gender- matched) over a time period ranging from October 2002 until December 2008 at the outpatient clinic of the Leiden University Medical Center (LUMC), the Netherlands. For the validation set, serum samples were collected from 39 pancreatic cancer patients who were selected candidates for curative surgery and 75 healthy (age- and gender- matched) controls over the time period ranging from January 2009 until July 2010. The available samples from both calibration and validation sets were distributed over three distinct MALDI-target plates as follows: the first plate contained 96 samples (59 controls vs. 27 cases) from the calibration set, the second plate contained 96 samples (60 controls vs. 36 cases) from the validation set and the last plate contained the remaining 63 samples from the calibration set (41 controls vs. 22 cases) as well as the remaining 18 samples from the validation set (15 controls vs. 3 cases). Cases and controls were randomly allocated to each distinct plate using the above design. Sample placement on the plates was also randomized between cases and controls using a randomized block design. Each sample was spotted in two replicates onto the MALDI-target plate and mass-analyzed by a MALDI-FTICR MS system which was optimized with regard to both sensitivity and resolving power. Each mass spectrum was obtained from the sum of 10 scans of 150 laser shots each and stored on a grid of 512 K data points, covering the mass/charge range from 1013 to 3700 Da. The output MALDI-FTICR profiles were exported as separate files for each individual, containing mass/charge (\textit{m/z}) values with their corresponding intensities.
\\
\indent Figure 1 shows an example of a mass spectrum of a case sample from the calibration set. A mass spectrum consists of peaks with a certain intensity (\textit{i.e.} height) distributed over a \textit{m/z}-axis generated from the detection of ionized molecules. In ultrahigh-resolution mass spectrometry, each species (such as a peptide) with a given elemental composition is detected as a cluster of isotopic peaks resulting from the distribution of naturally occurring elements. These isotopic peaks originate from ions with identical molecular formulas, but with different combinations of atoms containing additional neutrons. For example, ${}^{12}{\mathbf{C}}$, ${}^{13}{\mathbf{C}}$ and ${}^{14}{\mathbf{C}}$ are three isotopes of the element carbon with atomic masses 12, 13 and 14 mass units respectively. The atomic number of carbon is 6, which means that every carbon atom has 6 protons, so that the neutron numbers of these isotopes are 6, 7 and 8 respectively. In a mass spectrum, the successive isotopic peaks of a singly-charged species are approximately 1 Da apart and form a so-called isotopic cluster. With high-resolution devices, isotopic clusters of peaks can be completely resolved, whereas for low-resolution devices, the same clusters would appear as a single peak in the mass spectrum. Superimposed in Figure 1 we plot an isotopic cluster at position \textit{m/z} $2021.2$ .

\subsection{Identification Algorithm - Data preprocessing}
In this section we describe a procedure to preprocess the raw data and to translate the isotopic expression in the individual spectra into a condensed data set. The objective is to identify isotopic clusters and quantify their corresponding isotopic peaks. The algorithm we propose is particularly relevant in the context of large-scale clinical studies as it is simple and fast, and it relies solely on the known statistical properties of MALDI-FTICR MS data, in particular that successive isotopic peaks of a peptide molecule are commonly separated by 1 Da. Thus, it can be applied fast across many spectra from different patients. The identification algorithm finds \textit{m/z} positions of peaks which belong to isotopic clusters in a completely non-parametric fashion, thus avoiding any computationally intensive steps like model fitting to the observed spectra.
\\
\indent The preprocessing procedure consists of four main steps: detection of peaks, identification of isotopic clusters, alignment of the peaks within identified clusters and determination of peak intensities. In the first step, the procedure operates on every sample separately and finds peak locations of potential isotopes. In the second step, the procedure operates across all available samples and assigns the identified peak locations obtained from the first step to clusters via hierarchical clustering. In the third step the peaks within each cluster are aligned to represent the same biological peak within the same cluster. In the final step, the procedure operates on each individual spectrum, records and quantifies the peaks which are common with those from the consensus list of the previous step.
\\
\indent To identify peak locations of isotopes \textit{for each sample}, we first select all \textit{m/z} values of which the corresponding intensity values exceed a certain threshold. The threshold was defined as the noise threshold from a visual inspection of a random sample of individuals' spectra. This selection removes on the one hand noisy signal and on the other reduces the number of data points in each sample. We retain the \textit{m/z} values above the specified threshold, resulting in a set of $g$ connected intervals which we denote by $(I_g)_{g=1,...,G}$. Within each interval $I_g$, we determine the \textit{m/z} value $x_g$ corresponding to the maximum intensity value. For each patient we thus have a list $(x_g)_{g=1,...G}$ of peak locations, and for each peak location $x_g$ we have a corresponding interval $I_g$. Of the resulting list we only retain all $x_g$ for which either $x_g+1$ or $x_g-1$ is also within another connected interval $I_{g'}$. As we except the proteome to express in isotopic clusters, this assures that the list contains all $x_g$ values which are likely to correspond to successive peak locations within isotopic clusters as opposed to single (noisy) peaks along the mass spectrum. In this way we obtain for each $i^{th}$ patient a list of $x_{ig}$ values, corresponding to locations of consecutive isotopic peaks.
\\
\indent To identify clusters of peaks belonging to the same isotope clusters \textit{across individuals}, we consider the following basic relation:
\begin{equation}
 x_{ig}\sim x_{i'g'} \quad \text{if for some} \quad k \in \{-1,0,1\} : |x_{ig}-(x_{i'g'}-k)|<\delta,
\end{equation}
on the union of all peak lists across all patients, and define clusters as connected components  when viewing this relation as a graph on this set. Computationally, we approach this by applying single linkage clustering, regarding the above relation as a similarity measure with the value $1$ when the relation is present and $0$ otherwise, making thus use of the well known relation between single linkage clustering and the minimum spanning tree.
\\
\indent Within each cluster we define a secondary relation on $x_{ig}$ by:
\begin{equation}
 x_{ig}\sim x_{i'g'} \quad \text{if} \quad |x_{ig}-x_{i'g'}|<\delta'
\end{equation}
and group peaks, again by considering the connected components of this relation. We choose $\delta'$ so that the minimal distance between the resulting intervals is larger than $0.5$. We finally extract the centroid of each peak - the average value of $x_{ig}$ for each sub-cluster in each cluster - to represent the ``consensus'' position for that peak within that cluster across all spectra.
\\
\indent The above procedure results in a list of clusters $(C_q)_{q=1,...,Q}$ and a list of ordered peak positions $(x_{qj})_{j=1,...J_q}$ within each cluster with distance roughly 1 Da between neighbouring elements. Given the list of common peak positions from the clustering step, we record \textit{for each individual spectrum} all the intervals $I$ from the detection step which include one of the $x_{qj}$ values of our list. We quantify the peaks at the identified locations $x_{qj}$ in the individual spectra by integrating the area under the intensity curve over $I$. Hence, for peak $j$ in cluster $q$ and patient $i$ we can either find such an interval in the patients' list of intervals (an event which we denote in the following by $\delta_{iqj}=1$) and have a corresponding intensity level $y_{iqj}$, or fail to find such an interval ($\delta_{iqj}=0$). In the following, we call a peak $j$ undetectable in cluster $q$ and patient $i$, if $\delta_{iqj}=0$. We call a cluster $q$ undetectable in patient $i$ if $\delta_{iqj}=0$ for all peaks $j$ of the cluster. Each of these possible situations are shown in Figure 2, where we plot an identified isotopic cluster at 2553 m/z for three different samples. While the isotopic cluster is completely observed in the first sample, the very same cluster is only partly observed in the second sample and completely non-observed in the third sample.
\\
Ideally, each identified isotopic cluster corresponds to one unique isotopic distribution. Nevertheless, since the algorithm is using the distance along the \textit{m/z} axis to identify isotopic clusters, we must allow for the risk of joining two distinct isotopes into the same cluster. This issue may arise for instance in the situation where the distance between two consecutive peaks belonging to different isotope distributions is approximately $1$ Da. Hence, we suggest to inspect the identified clusters for signs of bimodal patterns prior to any further analysis.

\subsection{Summary measures}
In this section we present a non-exhaustive overview of simple measures to summarize the proteomic expression, detected for one isotopic cluster in one patient, in a meaningful way. In particular, we consider summary measures which aim to estimate the intensity level or the shape of the observed isotopic cluster pattern.

\paragraph*{Notation}
\indent In the following we omit subscript $q$ and consider one cluster consisting of $J$ peaks, where $J$ is cluster-specific. While some of our summary measures will be directly based on the indicators $\delta_{ij}$, most of them will be based on assuming $y_{ij}$ is known for all patients. Hence, if $\delta_{ij}=0$, we code $y_{ij}$ as the area under the intensity curve in a systematic interval around $x_j$ with length corresponding to the typical peak width in a specific \textit{m/z} range. The typical peak width is estimated as the average length of the connected intervals including $x_j$. Moreover, we denote with $l_{ij}=\log y_{ij}$ the log-transformed peak intensity and with $\bar{l}_j := \frac{1}{n}\sum_{\substack{i}} l_{ij}$ the typical/expected peak intensity, defined as the average of the log intensities across all samples.

\subsubsection{Intensity summary measures}

\paragraph*{Binary coding method}
\indent We consider that the isotopic response in the pancreatic cancer data is often partly or entirely undetectable across and within patients. A simple procedure to summarize the isotopic expression is to recode the original data into present/absent type of data, at either the peak or the isotopic cluster level. We investigate if the present/absent status is informative at the peak level, by ignoring $y_{ij}$ and using the binary indicator $\delta_{ij}$, with $1$ if a peak is detected and $0$ if a peak is undetectable. To investigate if the present/absent status is informative at the cluster level, we follow a similar approach and recode the data to $1$ if a cluster is completely or partly detected and $0$ if a cluster is completely undetectable. We consider the binary method as a reduced version of a data-summary approach, in the context of which we reduce the information in the original intensity measures to present/absent type of measures rather than considering directly the absolute intensities themselves.

\paragraph*{Ranking method}
\indent An extension to the binary method, familiar in biomedical research, is an ordinal ranking method (\citealp{dennis}). This method incorporates information on the intensity level, by using the ranks of the data in order to preserve the information on the ordering of the detected intensities. The intensities corresponding to the same biological peak are ranked separately. Ranks of data tied to the same value are themselves tied to a value equal to the average of the ranks they would have had if there had been no ties. After ranking the intensities across samples, the observed data are replaced with their ranks. To have one unique value per cluster, we define our summary measure as the sum of the resulting ranks $R_1,...,R_J$ within each cluster for each patient, given by $SR=\sum_{\substack{j}}R_{j}$. Alternatively, we may consider as our summary measure the rank at the cluster maximum, given by $MR=R_{j(m)}$, where $j(m)$ denotes the position where $\bar{l}_j$ is maximum and hence fixed across patients.

\paragraph*{Quantile method}
\indent An other method which uses information across all available samples is quantile normalization. This method is commonly used in genomics research for microarray data preprocessing (\citealp{quant}). We propose to use the quantile method to make the distribution of peak intensities across different samples comparable. More specifically, each sample is given the same distribution by estimating the mean quantile across samples and substituting this as the value of the intensity in the original data set. To achieve this, we first assign an index to each intensity value within each sample. Subsequently, we sort all samples by their intensities and substitute each sorted intensity in each sample with the mean of the sorted intensities across samples. The normalized intensity for each peak is derived by restoring the original order of the assigned indexes for each sample. We may consider as our summary measure, either the sum of the resulting normalized intensities $Q_1,...,Q_J$ within each cluster for each patient, given by $SQ=\sum_{\substack{j}}Q_{j}$, or the normalized intensity at the cluster maximum, given by $MQ=Q_{j(m)}$.

\paragraph*{Sum of log-intensities}
\indent A simple measure to summarize the isotopic expression while using the complete information on the intensity level across samples is to sum all peak intensities within the same cluster for each patient. We consider the log scale instead of the original scale which is known to be an appropriate scale for mass spectrometry data (\citealp{log}). We define our intensity measure for each patient and each cluster as the sum of log-transformed peak intensities, given by $SL=\sum_{\substack{j}}l_{j}$.

\paragraph*{Maximum log-intensity}
\indent Alternative to using the sum of all (log-transformed) intensities as summary measure is to use only the intensity at the cluster maximum. We define the intensity at the cluster maximum for each cluster and each patient as $ML=l_{j(m)}$.

\subsubsection{Shape summary measures}
\indent In what follows, we propose different summary measures which aim to estimate the shape of the observed isotopic cluster pattern for each patient. Note that all our shape summary measures are invariant under adding a constant to the log intensities or multiplying all intensities of a isotopic cluster in a patient with a factor. Hence, on a conceptual level, they do not carry any information on the intensity level.

\paragraph{Defining residuals suitable for shape investigation}
\indent Alternatively to considering the observed isotopic cluster pattern directly, we may investigate the deviation of that observed pattern from the typical pattern. We define the typical pattern - as before -  as the average of the log-intensities across samples given by: $\bar{l}_j :=  \frac{1}{n} \sum_{\substack{i}} l_{j_i}$. We define the corresponding typical pattern for the absolute intensities by the back transformation $\bar{y}_j :=  \exp{(\bar{l}_j)}$. To evaluate the deviation between $y_1,...,y_J$ and $\bar{y}_1,...,\bar{y}_J$, while accounting for the fact that absolute peak intensities can vary greatly across isotopes and between samples, we scale up $\bar{y}_1,...,\bar{y}_J$ by a factor $\beta$. To determine $\beta$, we first consider the relationship $\alpha=\log \beta$. We choose $\alpha$ to minimize $\sum_{\substack{i}}(l_{j_i}-\bar{l}_{j_i}-\alpha)^2$, resulting in $\alpha=\frac{1}{J}\sum\limits_{j=1}^J(l_j-\bar{l}_j)$ and $\beta=\exp(\alpha)=\exp\Big(\frac{1}{J}\Big)\prod\limits_{j=1}^J\frac{y_j}{\bar{y}_j} = \left(\prod\limits_{j=1}^J\frac{y_j}{\bar{y}_j}\right)^{ \frac{1}{J} }$, the geometric mean of $\frac{y_j}{\bar{y}_j}$. In this way, we obtain the residuals $s_j=l_j-\bar{l}_j-\alpha$ and $r_j=\exp(s_j)=\frac{y_j}{\bar{y}_j\beta}$. In the following, we describe summary measures based on shape as a function of $x_j$, $j=1,...,J$ where $x_j$ may be either one of the choices $y_j$, $l_j$, $r_j$ or $s_j$.

\paragraph*{Pairwise ratios or differences}
\indent A simple and straightforward approach to evaluate shape information is to look at the pairwise ratios $Q_{jj'}=\frac{x_j}{x_{j'}}$. We can either use all pairwise ratios as input or just a selection. For instance, we can use the consecutive ratios $Q_{1 2},Q_{2 3},...,Q_{J-1 J}$ or choose $Q_{j(1)j(2)}$ with $j(1)$ the position with maximum $\bar{y}_j$ and $j(2)$ the position with second maximum $\bar{y}_j$. Pairwise ratios can be calculated for either $i$ or $r$. The analogous measure for $l$ or $r$ is to use the pairwise differences $\Delta_{jj'}=x_j-x_{j'}$.

\paragraph*{Peak location and related measures}
\indent Information on the shape can also be extracted by examining the position of the mode of the cluster pattern distribution or the relative position of all values in the ordering as to the position of the mode. Let $R_1,...,R_J$ be the ranks of $x_1,...,x_J$. The vector $R_1,...,R_J$ is itself a measure that provides information about the shape. Let now $R^{-1}(1),...,R^{-1}(J)$ be the inverse ranks, such that $R(R^{-1}(j))=j$. Then $R^{-1}(J)$ is the location of the peak (mode) of the cluster pattern distribution which is itself a measure which contains information on the shape. We may also use the entire vector $R^{-1}(J)$, $R^{-1}(J-1),...,R^{-1}(1)$ as input or a subset $R^{-1}(J)$, $R^{-1}(J-1),...,R^{-1}(J-k)$. These shape measures can be applied to either one of $y$, $j$, $r$ or $s$.

\paragraph*{Distributional shape}
An isotopic cluster as shown in Figure 1 bears resemblance to a histogram. Therefore, we may regard the values $x_1,...,x_J$ as heights of this histogram, for a distribution on the values $1,...,J$, and consider some characteristics of this distribution. To do that we first define the center of gravity of a distribution as
\[cg:=\sum_{\substack{j}}j p_j\]
where $p_j:=\frac{x_j}{\sum_{\substack{j}}x_j}$. The center of gravity is already a measure which carries information on the shape of the distribution and can therefore itself be used as a summary measure. Furthermore, we consider the following distributional shape measures: 1) spread, which measures the amount of variation or dispersion of the distribution, 2) skewness, which measures the asymmetry of the distribution and 3) kurtosis, which measures the peakedness of the distribution, based on the following definitions:
\[spread:=\sqrt{\sum_{\substack{j}}(j-cg)^2 p_j}\]
\[skewness:=\frac{\sum_{\substack{j}}(j-cg)^3 p_j}{\Big(\sum_{\substack{j}}(j-cg)^2 p_j\Big)^{3/2}}\]
\[kurtosis:=\frac{\sum_{\substack{j}}(j-cg)^4 p_j}{\Big(\sum_{\substack{j}}(j-cg)^2 p_j\Big)^{2}}\]
Note that these definitions can be applied to any distribution defined on at least two points and are hence applicable to all clusters. Since $x_1,...,x_J$ need to be positive, the above measures can be derived in principle either for $y$ or $r$.

\paragraph*{Tendencies in the residuals}
\indent An alternative way to extract information on the isotopic cluster pattern is to examine the behavior of the residuals along the isotopic path. For instance we may investigate possible tendencies for the residuals to increase or decrease with $j$, or tendencies for the residuals to be higher in the middle or higher at the borders/tails of the isotopic cluster pattern distribution.
To conduct these investigations we fit the orthogonal polynomial
\[x_j=\alpha+\beta\big(j-\bar{J}\big)+\gamma\big(j-\bar{J}\big)^2+\varepsilon_j \]
We choose $\alpha$, $\beta$ and $\gamma$ to minimize
\[\sum_{\substack{j}}\bigg(x_j-\Big(\alpha+\beta\big(j-\bar{J}\big)+\gamma\big(j-\bar{J}\big)^2\Big)\bigg)^2\]
We use the estimates of $\beta$ and $\gamma$ as our summary measure. This approach requires $J \geq 3$ and can be applied to either one of $r$ or $s$.

\section{Application and analysis}

\subsection{Identification algorithm implementation}

\indent We applied the identification algorithm to the individual raw spectra, to obtain the condensed data set used to derive the summary measures. The algorithm was implemented in Matlab using programmes developed by the authors. All functions used can be found in the supplementary file (Additional file 1). The concrete values of our algorithm thresholds are specific for our data and were chosen on the basis of visual inspection of the individual spectra and random testing. In particular, the threshold introduced in the first step of the algorithm, to identify peak locations, was set equal to $0.8 \times 10^6$, which corresponds to the estimated average spectra noise threshold. We choose the distance $\delta$ of our basic relation in the cluster identification step to be equal to $0.08$ which was proved to be a fairly reasonable choice. This distance is an important adjusting parameter in our procedure and one should try to vary it and examine the results visually. Finally, we choose $\delta'$ to be equal to $0.1$, which was found to be the proper value for ensuring a minimal distance of $0.5$ between the resulting intervals representing consecutive peaks.
\\
\indent From the resulting list of clusters, we discarded those which were more likely to correspond to joined isotopes. We did so by removing the clusters found with more than $8$ peaks and more than $1$ local maxima. This resulted in a total of $2717$ identified isotopic clusters and $8080$ isotopic peaks.  The average size of the identified clusters is $3.3$ peaks, while the minimum and maximum sizes are 2 and 10 peaks respectively. Of the identified clusters and peaks, only a small number was identifiable in all samples. The resulting data set contained thus a large proportion of undetectable peaks (85 \%). Figure 3 shows a graphical representation of the magnitude of incomplete isotopic expression, at the cluster and the peak level separately.

\subsection{Model fitting and results}

\indent To evaluate the impact of the different summary measures on class calibration and prediction, we fit a model each time using one of the summary measures as input variables and we evaluate the predictive performance and predictive accuracy of each such fit. We should mention at this point that shape summary measures were derived solely for detected or partly detected clusters. In case an entire cluster is undetectable in a sample, we impute the average value of the derived shape measure across all samples, prior to fitting the model.
\\
\indent We choose to use ridge logistic regression as our classifier, which is a well established method in applied sciences and classification literature and proved effective in high dimensions (\citealp{ridge}). This method is commonly used in situations where the number of covariates exceeds the number of observations and/or there are high correlations between them. Ridge logistic regression shrinks the regression coefficients towards zero by imposing a penalty on their size.
\\
\indent Our procedure consists of two phases. The first phase can be considered as a type of internal validation. During this phase, we re-define the calibration-validation structure, using random splitting, in order to obtain more robust results. We randomly assign each observation to one of two sets as follows: we begin by selecting a case from the overall set of samples and assigning it to one of two sets. Since the case-control ratio in each plate was $1/2$, for each selected case, two controls from the same plate are assigned to the same set as that case. In this way we preserve the plate information. This random splitting is repeated $10$ times. Classification results across repetitions are finally averaged to obtain more stable estimates. To evaluate and compare the predictive ability of each model using each summary measure, we begin with a double-cross validatory approach, within the re-defined calibration set, consisting of two nested loops (\citealp{bart,stone}). The inner loop is used to determine the optimal tuning parameter, while the outer loop is used to estimate the predictive performance of the approach across all observations, by applying the chosen optimized rules from the inner loop to the left-out datum. To evaluate performance of the final diagnostic rule on the set aside validation set, we perform leave-one-out cross validation to the calibration set to estimate the optimal tuning parameter and we apply the resulting rule to the separate validation set. All the computations were carried out in Matlab using programs written by the authors.
\\
\indent Table 1 shows estimates of the predictive performance measures of the ridge logistic model fitted to each intensity measure, for the calibration and validation sets, together with their standard errors. For each model we calculate the error-rate and the area under the ROC curve (AUC). To evaluate the accuracy of each fit, we also calculate the Brier score and the deviance, defined as
\begin{eqnarray*}%
Brier\;score &=& \frac{1}{n}\sum\limits_{i=1}^n (\hat{p}_i-c_i)^2
\end{eqnarray*}
\begin{eqnarray*}%\label{eqexpmuts}
Deviance &=& -2\sum\limits_{i=1}^n c_i\log{\hat{p}_i}+(1-c_i)(\log{(1-\hat{p}_i})
\nonumber\\
&=&-2\sum\limits_{i=1}^n \log(1-|\hat{p}_i-c_i|)
\end{eqnarray*}
where $\hat{p}_i$ is the estimated probability of being a case for the $i^{th}$ individual, $c_i$ is the known health status of that individual and $n$ is the total sample size. For class assignments, we use a threshold of $0.5$ and thus assign an observation as a disease case if the estimated class probability $\hat{p}_i$ is greater than $0.5$ and as a control otherwise.
\\
\indent Classification results, using the binary indicator as our only predictor, suggest that the present/absent status, either at the peak (BP) or at the cluster (BC) level, is highly informative with regards to the class outcome. Using the sum of intensity ranks (SR) as summary measure seems to be recovering information on top of the present/absent information, which results in improved predictions compared to BP and BC. Looking at the overall results, we find best classification performance achieved when using the sum of quantile normalized intensities (SQ) or the sum of log-intensities (SL) as summary measures. The ranking method and the quantile normalization share the characteristic that summary measures on one observation are based on borrowing information across all available observations. Quantile normalization is a widely used pre-processing procedure applied to genomic data in order to remove technological noise and has been proved to give favorable results in differential expression analysis. Moreover, we observe that using solely the values at the cluster maximum, to summarize the entire isotopic expression, does not capture the entire information in the overall intensity level. This becomes clear looking at validated classification results based on the max log-intensity (ML), the max rank intensity (MR) and the max quantile normalized intensity (MQ). These results confirm that the above measures are less powerful, in terms of predictive ability, compare to summary measures based on cumulative cluster information, such as the sum of log-intensities (SL), the sum of rank intensities(SR) or the sum of quantile normalized intensities (SQ).
\\
\indent In Table 2 we report estimates of the predictive performance measures based on shape summary measures, calculated on the vector of residuals $r$. We choose to present results based on $r$, on the one hand, because all the proposed shape measures are applicable to $r$, and on the other, because shape measures calculated on $r$ provided slightly better results compared to shape measures calculated on $y$, $l$ or $s$. Results based on a collection of shape measures calculated on $y$, $l$ and $s$ can be found in the supplementary file (Additional file 2). The conclusion which can be drawn from this data analysis is that, on average, the observed isotopic cluster patterns differ in shape between healthy and diseased individuals. In particular, the center of gravity of the distribution of the residuals (cg) and the spread of the residuals ($spread$) which measures the deviation from the expected isotopic pattern conditional on relative position $j$, are the shape measures which provided the best classification results with error rates $0.177$ and $0.175$ respectively. Predictive performance measures using the vector of the ranks of the residuals ($\mathbf{R}$) or the vector of the inverse ranks of the residuals ($\mathbf{R^{-1}}$) are fairly close to results using the center of gravity or the spread of the distribution of residuals, suggesting that the overall pattern of deviation varies between the two groups. Comparing the results obtained using intensity measures to the ones obtained using shape measures, we can conclude that information on the intensity level has a higher predictive potential, as compared to information on the isotopic cluster shape. Nevertheless, classification results using the center of gravity or the spread of the distribution of residuals in particular, are fairly close to those obtained using the majority of intensity summary measures.
\\
\indent The second phase of our evaluation procedure constitutes a type of external validation. During this second phase, we build the prediction model using each summary measure on the calibration set, as defined in the original experiment, and evaluate the resulting rule on the separate validation set, following the same cross-validatory approach as in the first phase. Classification results based on intensity summary measures for the original calibration and validation sets are presented in Table 3. The ranking of the intensity measures, with respect to their predictive performance, is in agreement with the average ranking using the redefined calibration-validation structure. Classification results based on a collection of shape summary measures are presented in Table 4 in which we observe again a ranking of the measures similar to the internal calibration.
\\
\indent A remark worth mentioning at this point lies in the fact that classification results for some summary measures, go into an unexpected direction. That is, we occasionally observe better discrimination in the validation set than in the calibration set. A possible explanation of this phenomenon is the fact that the samples for the calibration set were collected in an earlier period than the samples for the validation set and were stored long before they were actually processed by the mass spectrometer. This resulted in a considerable aging of the calibration samples on which the discriminating rule was based. In spite of this aging, it looks like the ridge model was able to construct a reasonable rule, which worked better on samples not affected by the aging.
\\
\indent To evaluate if the proposed shape summary measures have predictive ability in their own right and are independent of any information related to the overall intensity level, we estimate the added value of shape over intensity. We regress, for each cluster separately, shape on intensity and use the residuals of that fit to predict the class outcome. For flexibility, we choose to use cubic splines to model the relationship between shape and intensity. The key idea of this approach is to discard the shared information between shape and intensity and use the remaining residual information, which is independent of the overall intensity level, to assess the predictive ability of shape. Putting the isolated information of shape into our prediction model resulted in an error rate of 0.27 and 0.28 (AUC of 0.77 and 0.78), for the calibration and validation sets respectively, which are smaller than the error-rates (larger than the AUCs) we would expect to occur by chance. This outcome suggests that the shape of the observed isotopic cluster pattern holds information, with regard to classification, independent of the overall intensity level.
\\
\indent We finally investigate whether we can enhance the predictive performance of our diagnostic rule by integrating both types of information. We choose the sum of log-intensities as the intensity measure to be combined with a shape measure. Combining SL and $cg$ resulted in an average error-rate of $0.097$ and $0.107$ for the calibration and validation sets respectively. Hence, the combination of these measures does not result in prediction improvement, despite of our previous finding that the two different types of information are ``conceptually independent'' and have predictive power in their own right. Similar outcomes occurred when combing SL and $\mathbf{R}$, SL and $\mathbf{R^{-1}}$, SL and $spread$, SL and $skew$ as well as SL and $kurt$. These results suggest that the additional value of shape information is actually so small that ridge regression cannot make use of it when confronted with the task of selecting information from both intensity and shape measures.

\section{Discussion}

\indent Here we have presented a statistical report in which we considered the problem of summarizing and analyzing high-throughput mass spectrometry data in an appropriate and meaningful way. To this end, a fast and simple procedure to preprocess the individual spectra prior to summarizing and analyzing the acquired data was proposed. We have presented simple summary measures which can be used as input variables for the construction of prediction rules for disease status allocation of future patients. These summary measures were based on using information on the intensity level or the shape of the observed isotopic cluster pattern. An extensive evaluation of the proposed measures was performed with respect to their predictive ability. Results using these measures indicated that both the intensity level and the isotopic cluster shape are related to the class outcome and can be used to predict the presence or absence of the disease. It was noted that summary measures based on shape were less informative in terms of predictive potential when compared to summary measures based on intensity level.
\\
\indent Previously, various algorithms have been reported to de-isotope and preprocess high-resolution mass spectral data (\citealp{averagine,roc,palm,valk,yuri}). In particular the idea of using the 1 mass unit distance to identify isotopes can be found in \citet{horn} and \citet{park}. The approach described in Horn et al. is applicable to a single spectrum and it is based on determining best fitting local models. This computationally intensive task is avoided in our algorithm by using information across all potential isotopic peaks across all patients. Hence the algorithm we propose can be applied fast across many spectra from different patients. This renders the approach more suitable for clinical applications. Since our algorithm is based solely on the distance along the m/z-axis, it has the limitation that (occasionally) distinct isotope distributions can be joined into the same cluster. The proposed algorithm is specific to this particular type of data and can be applied to spectra containing singly charged ions resulting in a distance of 1 Da between successive isotopic peaks. As mentioned in the Introduction section, electrospray ionization (ESI) yields multiply charged ions, resulting in a reduced distance inversely proportional to the number of charges on the molecule. Applying the identification algorithm to spectra containing multiply charged ions therefore requires appropriate adjustments.
\\
\indent A crucial component in our data was the large number of (unidentified) peaks (or clusters), which is a common outcome in many FTICR-MS clinical applications. For this reason, we decided to start our investigations by evaluating whether the presence/absence mechanisms of the isotopic expression are associated with the class outcome. Results from this investigation showed that the binary indicator, either at the peak or at the cluster level, is itself a good predictor of the case-control outcome, suggesting the present/absent patterns are different between the two groups. Incorporating additional information on the intensity level, by using for instance the ranks of the data, in order to preserve the information on the ordering of the detected intensities provided improved estimates compared to using information on the presence/absence status only.
\\
\indent While intensity measures as discussed in this paper can vary in their complexity, they were all proved to be highly informative with regards to the case-control outcome. Intensity measures were based on the log transformed peak intensities as this is considered an appropriate scale for high-throughput mass spectrometry data which often cover a large range of values, spanning several orders of magnitude. Nevertheless, intensity measures presented in this paper can be derived using either log or raw intensities.
\\
\indent A second objective of this research was to investigate whether there is additional information in the shape of the observed isotopic cluster pattern, related to the case-control outcome which is independent of information in the intensity level. We proposed several measures to carry out these investigations, applied either to the observed isotopic cluster pattern directly or applied to the residuals measuring the deviation of that observed pattern from the expected one. In contrast to intensity summary measures, we consider shape measures based both on log and raw peak intensities, since the latter defines the original shape of the observed isotopic cluster pattern. Results based on these measures indicated that the shape of the observed isotopic cluster pattern is predictive of the class outcome, yet less informative as compared to the overall intensity level. While this finding may appear odd at first sight, since isotopes could be hypothesised to have a fixed shape distribution, the observed shape effect may be explained by the occurrence of overlapping compounds within one isotopic cluster (\citealp{averagine,dirk}). For instance, in case of two overlapping compounds differing by exactly 1 Da (measured as one m/z-unit difference) we actually analyse by our approach the sum of two isotopic distributions. We emphasize at this point that it is very unlikely that the observed shape differences are the result of the measurement procedure itself, since both cases and controls samples were randomly allocated within the plates. However, linking biological explanations to the observed shape effect is/must be a topic of future research.
\\
\indent In order to understand the true value of the derived shape measures on top of the intensity level, we isolated the information not shared between shape and intensity and use this to predict the class outcome. Results indicated that the remaining information, after removing the one which was shared, has still predictive power in its own right. Nevertheless, the additional value of shape information was insufficient as to allow for improved predictions when being combined with information on the intensity level. As an alternative to combining intensity and shape measures into the same model, we explored the predictive potential of combining shape and intensity, by including all single peaks into the ridge model, allowing thus the model to pick up the most interesting combinations. The outcome of this procedure was similar to the one obtained when combining a shape with an intensity measure, which did not provide improved predictions.

%% BibTeX support
\bibliographystyle{DeGruyter}
%\bibliography{ref_kakourou}
\bibliography{Reference pdf}

%\begin{comment}
\clearpage

\begin{landscape}
\begin{table}[!h]
%\resizebox{\linewidth}{!}{% Resize table to fit within \linewidth horizontally
%\parbox{16.5cm}{\caption{Averaged ($\pm$SE) classification results based on intensity measures.}}
\caption{Averaged ($\pm$SE) classification results based on intensity measures for re-defined calibration and validation sets.}
\begin{center}
\resizebox{\columnwidth}{!}{%
\begin{tabular}{lllllllll}
\toprule
&\multicolumn{8}{c}{Calibration set (re-sampled)}  \\
\cmidrule(r){2-9}
& BP&BC&SR&MR&SQ&MQ&SL&ML \\
\midrule
Error-rate  & 0.133  (0.008)& 0.150 (0.009)  & 0.124 (0.007) & 0.143 (0.005) & 0.097 (0.005)& 0.102  (0.006) & 0.095 (0.006) & 0.108 (0.006) \\
Brier score & 0.094  (0.005)& 0.109 (0.005)  & 0.090 (0.004) & 0.107 (0.004) & 0.078 (0.004)& 0.077  (0.004) & 0.079 (0.003) & 0.080 (0.003) \\
Deviance    & 106.32 (3.472)& 120.91(4.132)  & 102.28 (3.61) & 117.35  (2.96) & 91.09 (4.01)& 80.05 (10.70) & 91.28 (3.22) & 91.64 (2.27)\\
AUC         & 0.931  (0.004)& 0.909 (0.007)  & 0.934 (0.004) & 0.921 (0.004) & 0.949 (0.003)& 0.955  (0.003) & 0.948 (0.003) & 0.951 (0.002) \\
\cmidrule(r){1-9}
&\multicolumn{8}{c}{Validation set (re-sampled)}  \\
\cmidrule(r){2-9}
& BP&BC&SR&MR&SQ&MQ&SL&ML \\
\midrule
Error-rate  & 0.144 (0.017)  & 0.162 (0.019) & 0.117 (0.005)& 0.147 (0.005)& 0.111 (0.005)& 0.132 (0.011) & 0.105 (0.005)& 0.120 (0.010)\\
Brier score & 0.105 (0.008)  & 0.118 (0.009) & 0.097 (0.003)& 0.114 (0.003)& 0.088 (0.003)& 0.096 (0.004) & 0.089 (0.002)& 0.096 (0.002)\\
Deviance    & 58.07 (3.637)  & 63.83 (3.690) & 53.08 (1.21)&  61.51 (1.98)& 50.38 (2.22)& 52.25 (3.09) & 50.29 (1.79)& 52.49 (2.29)\\
AUC   & 0.912 (0.013)  & 0.888 (0.016) & 0.927 (0.006)& 0.907 (0.007)& 0.938 (0.006)& 0.937 (0.007) & 0.935 (0.007)& 0.934 (0.006)\\
\bottomrule
\end{tabular}}
 \end{center}
\parbox{20cm}{BP, binary peak indicator; BC, binary cluster indicator; SR, sum of rank intensities; MR, max rank intensity; SQ, sum of quantile normalized intensities; MQ, max quantile normalized intensity; SL, sum of log-intensities; ML, max log-intensity.}
\label{Table 2}
\end{table}
\end{landscape}

%\clearpage
%\newpage
\begin{landscape}
\begin{table}[h!]
\caption{Averaged ($\pm$SE) classification results based on shape measures (calculated on raw intensity residuals) for re-defined calibration and validation sets.}
\begin{center}
\resizebox{\columnwidth}{!}{%
\begin{tabular}{lllllllllll}
\toprule
&\multicolumn{10}{c}{Calibration set (re-sampled)}  \\
\cmidrule(r){2-11}
& $\mathbf{Q}$&$Q_{j(1)j(2)}$&$\mathbf{R}$&$\mathbf{R^{-1}}$&$R^{-1}(J)$&$cg$&$spread$&$skew$&$kurt$&$(\beta,\gamma)$ \\
\midrule
Error-rate  & 0.263  (0.014)& 0.266   (0.009)& 0.183  (0.008)& 0.234  (0.007)& 0.258  (0.013) & 0.178  (0.006)& 0.177 (0.011)&0.204  (0.012) &0.235  (0.015)&0.277 \\
Brier score & 0.173  (0.005)& 0.182   (0.003)& 0.134  (0.004)& 0.163  (0.003)& 0.173  (0.006) & 0.126  (0.004)& 0.130 (0.006)&0.149  (0.006) &0.163  (0.006)&0.182 \\
Deviance    & 177.87 (4.356)& 185.45  (2.587)&142.20  (3.172)& 151.31 (15.47)& 175.98 (5.506)& 134.87 (4.268)& 140.2 (5.760)&146.92 (12.60)&169.34 (12.603)&181.95 \\
AUC         & 0.783  (0.014)& 0.743   (0.012)& 0.868  (0.006)& 0.800  (0.009)& 0.773  (0.020) & 0.878  (0.010)& 0.868 (0.014)&0.828  (0.014) &0.786  (0.015)&0.765 \\
\cmidrule(r){1-11}
&\multicolumn{10}{c}{Validation set (re-sampled)}  \\
\cmidrule(r){2-11}
& $\mathbf{Q}$&$Q_{j(1)j(2)}$&$\mathbf{R}$&$\mathbf{R^{-1}}$&$R^{-1}(J)$&cg&$spread$&$skew$&$kurt$&$(\beta,\gamma)$ \\
\midrule
Error-rate  & 0.254   (0.005)& 0.292  (0.008)& 0.201 (0.018)& 0.199  (0.010)& 0.237 (0.009)& 0.177 (0.009)& 0.175 (0.009) &0.203 (0.012)&0.232 (0.014)& 0.292\\
Brier score & 0.171   (0.005)& 0.188  (0.003)& 0.143 (0.006)& 0.149  (0.006)& 0.175 (0.005)& 0.123 (0.005)& 0.125 (0.006) &0.166 (0.005)&0.166 (0.004)& 0.185 \\
Deviance    & 86.44   (2.369)& 92.83  (0.989)& 73.84 (2.543)& 73.843 (3.184)& 89.02 (2.542)& 64.86 (2.662)& 66.43 (3.143) &76.60 (2.274)&83.97 (2.034) &109.95  \\
AUC         & 0.805   (0.020)& 0.745  (0.009)& 0.845 (0.012)& 0.834  (0.019)& 0.751 (0.017)& 0.883 (0.015)& 0.879 (0.018) &0.837 (0.015)&0.795 (0.017) &0.757 \\
\bottomrule
\end{tabular}}
\end{center}
\parbox{20cm}{ $\mathbf{Q}$, ratios of consecutive peak residuals; $Q_{j(1)j(2)}$, ratios of first and second maximum residuals; $\mathbf{R}$, vector of ranks of residuals; $\mathbf{R^{-1}}$, vector of inverse ranks of residuals; $R^{-1}(J)$, mode of distribution of residuals; cg, center of gravity of distribution of residuals; $spread$, spread; $skew$, skweness; $kurt$, kurtosis of distribution of residuals based on relative position $j$; $(\beta,\gamma)$, estimates of polynomial fit.}
%\label{Table 2}
\end{table}
\end{landscape}

\begin{table}[h!]
%\parbox{12cm}{\caption{Results based on intensity measures.}}
\caption{Results based on intensity measures for calibration and validation sets as defined in original experiment.}
\begin{center}
\begin{tabular}{lllllllll}
\toprule
&\multicolumn{8}{c}{Calibration set (as defined in original experiment)}  \\
\cmidrule(r){2-9}
& BP&BC&SR&MR&SQ&MQ&SL&ML \\
\midrule
Error-rate  & 0.126  & 0.133 & 0.120  & 0.147  & 0.133 & 0.133 & 0.133 &0.120   \\
Brier score & 0.097  & 0.105 & 0.093  & 0.114  & 0.088 & 0.096 & 0.088 &0.095   \\
Deviance    & 101.37 &110.07 & 91.05  & 113.92 & 91.16 & 95.65 & 90.12 &94.35   \\
AUC         & 0.907  & 0.892 & 0.936  & 0.888  & 0.936 & 0.925 & 0.934 &0.925   \\
\cmidrule(r){1-9}
&\multicolumn{8}{c}{Validation set (as defined in original experiment)}  \\
\cmidrule(r){2-9}
& BP&BC&SR&MR&SQ&MQ&SL&ML \\
\midrule
Error-rate  & 0.153 & 0.163 & 0.115 & 0.173 & 0.096  & 0.125 & 0.096  &0.125  \\
Brier score & 0.107 & 0.113 & 0.088 & 0.122 & 0.082  & 0.089 & 0.077  &0.086  \\
Deviance    & 72.59 & 80.41 & 63.49 & 80.53 & 58.93  & 60.99 & 53.33  &58.05  \\
AUC         & 0.942 & 0.925 & 0.945 & 0.926 & 0.964  & 0.971 & 0.973  &0.975  \\
\bottomrule
\end{tabular}
 \end{center}
\parbox{14cm}{BP, binary peak indicator; BC, binary cluster indicator; SR, sum of rank intensities; MR, max rank intensity; SQ, sum of quantile normalized intensities; MQ, max quantile normalized intensity; SL, sum of log-intensities; ML, max log-intensity.}
\label{Table 3}
\end{table}

\begin{table}[h!]
%\parbox{10cm}{\caption{Results based on shape measures.}}
\caption{Results based on shape measures (calculated on raw intensity residuals) for calibration and validation sets as defined in original experiment.}
\begin{center}
\begin{tabular}{llllllllll}
\toprule
&\multicolumn{6}{c}{Calibration set (as defined in original experiment)}  \\
\cmidrule(r){2-7}
&$\mathbf{R}$&$\mathbf{R^{-1}}$&cg&$spread$&$skew$&$kurt$\\
\midrule
Error-rate  & 0.226  & 0.220 & 0.160  & 0.180  & 0.213  & 0.273  \\
Brier score & 0.166  & 0.140 & 0.123  & 0.121  & 0.139  & 0.186  \\
Deviance    & 151.36 & 128.54& 120.72 & 120.02 & 128.36 & 169.07 \\
AUC         & 0.792  & 0.865 & 0.872  & 0.875  & 0.863  & 0.670  \\
\cmidrule(r){1-7}
&\multicolumn{6}{c}{Validation set (as defined in original experiment)}  \\
\cmidrule(r){2-7}
&$\mathbf{R}$&$\mathbf{R^{-1}}$&cg&$spread$&$skew$&$kurt$\\
\midrule
Error-rate  & 0.230 & 0.163 & 0.182 & 0.182 & 0.192  & 0.298  \\
Brier score & 0.161 & 0.139 & 0.120 & 0.120 & 0.148  & 0.198  \\
Deviance    & 102.32 & 91.12 & 78.76 & 78.69 & 97.21 & 121.07 \\
AUC         & 0.836 & 0.858 & 0.899 & 0.896 & 0.841  & 0.699  \\
\bottomrule
\end{tabular}
 \end{center}
\parbox{14cm}{$\mathbf{R}$, vector of ranks of residuals; $\mathbf{R^{-1}}$, vector of inverse ranks of residuals; cg, center of gravity of distribution of residuals; $spread$, spread; $skew$, skewness; $kurt$, kurtosis of distribution of residuals based on relative position $j$.}
\label{Table 4}
\end{table}

\clearpage

\begin{figure}
    \centering
      \includegraphics[height=10cm]{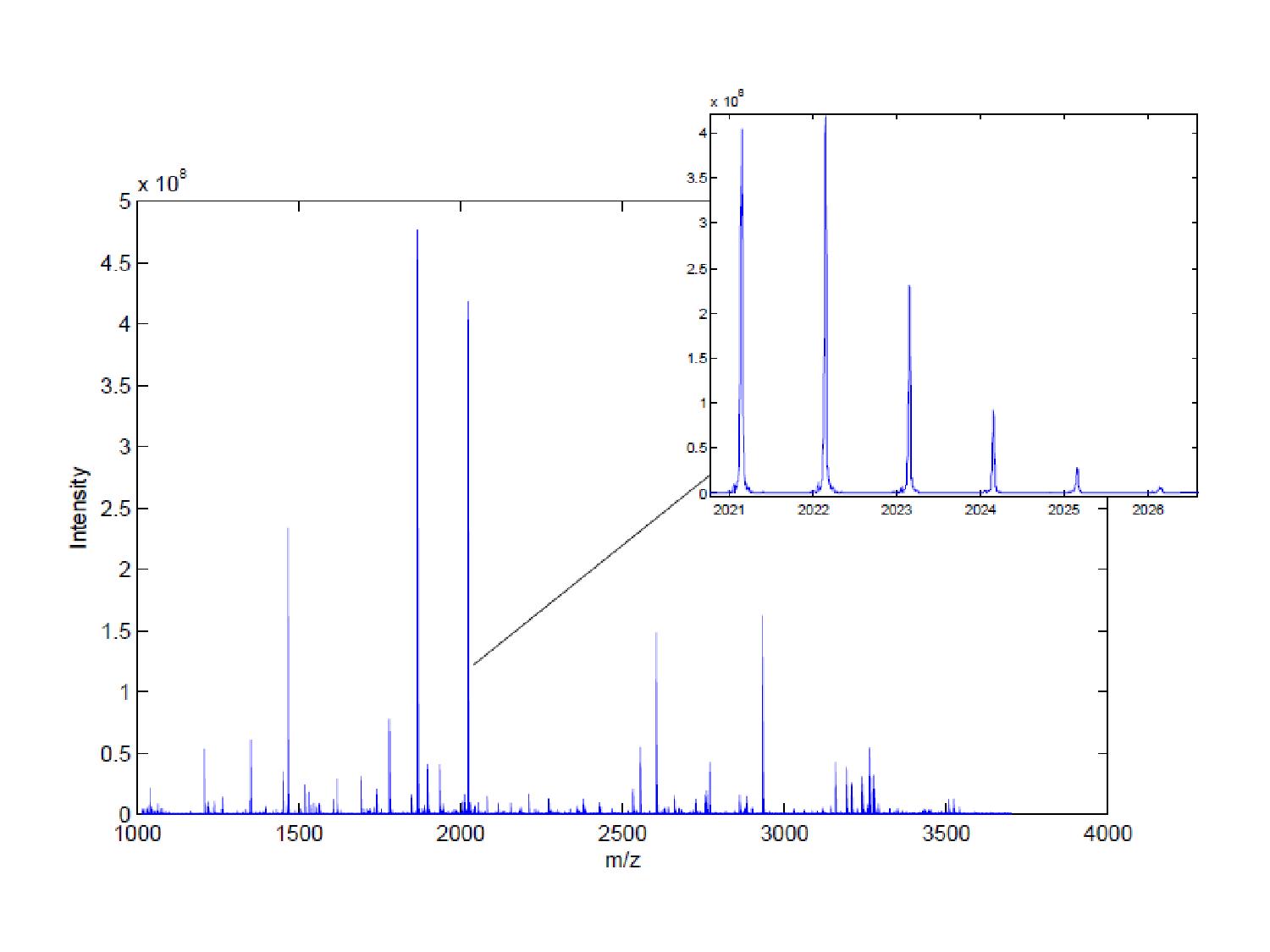}
      \caption{The mass spectrum of a single individual.
      Superimposed is shown an isotopic cluster at position \textit{m/z} 2021.2. }
\end{figure}

\begin{figure}
    \centering
      \includegraphics[height=13cm]{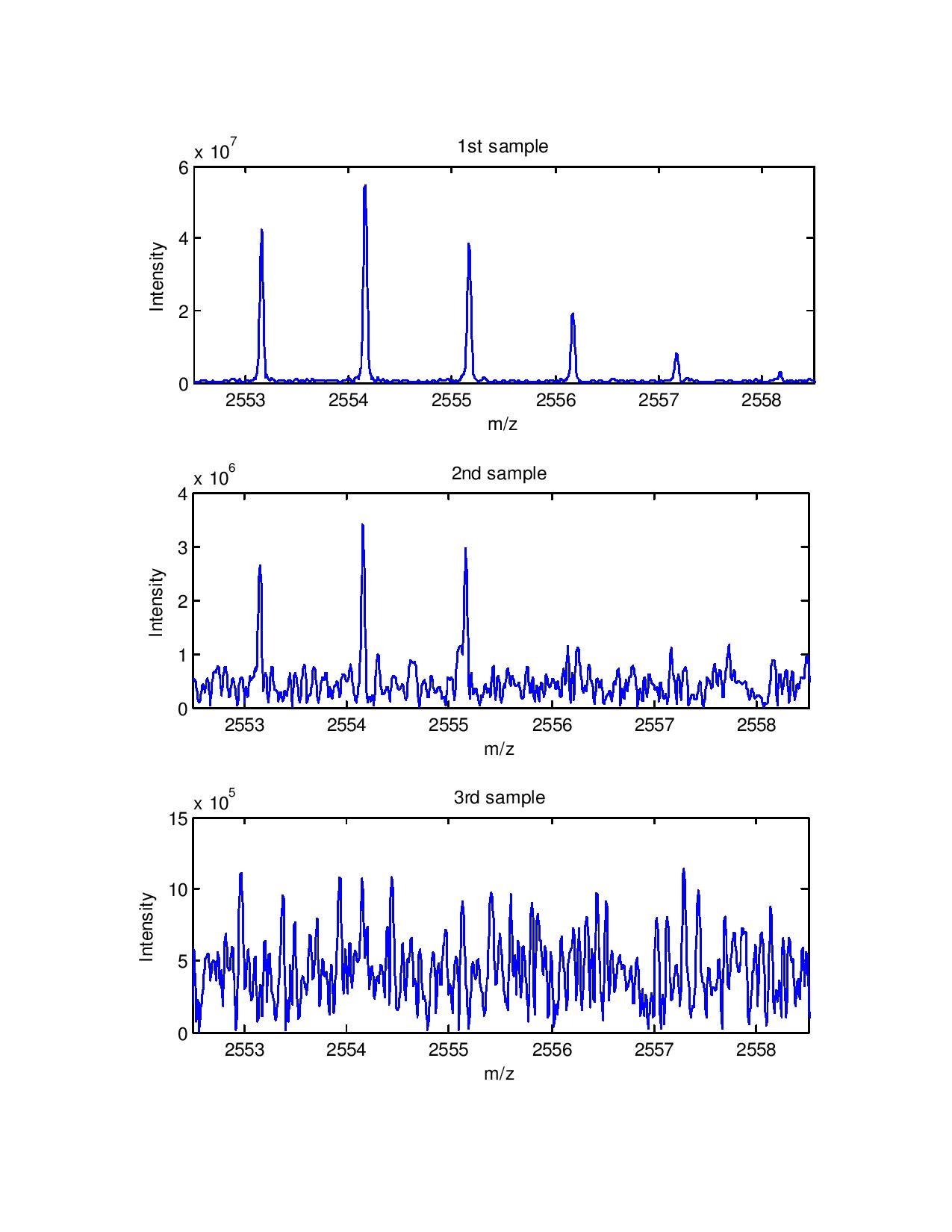}
      \caption{An identified isotopic cluster at 2553 m/z for three different samples.
     The isotopic cluster is completely observed in the first sample, only partly observed in the second sample and completely
non-observed in the third sample.}
      %\label{Figure 1}
\end{figure}

\begin{figure}
    \centering
      \includegraphics[height=9cm]{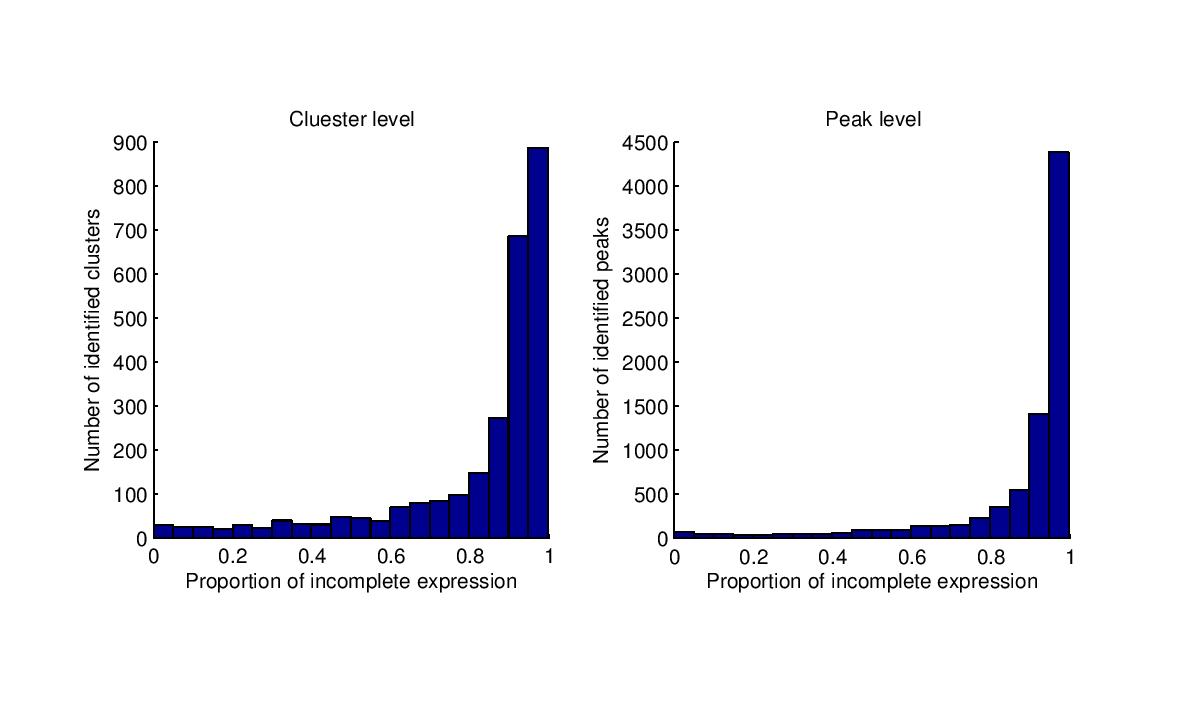}
      \caption{Presence/absence magnitude at cluster (left) and peak (right) level.
    Proportion of entirely undetectable clusters (left plot) and proportion of undetectable peaks (right plot).}
      %\label{Figure 1}
\end{figure}

%\end{comment}

\end{document}